\documentclass[twocolumn]{aastex6}
\usepackage{graphicx}
\usepackage{amssymb,amsfonts,amsmath,amstext,amsgen,amsopn,amsxtra,indentfirst,times}

\begin{document}

\title{Analytical Models of Exoplanetary Atmospheres. \\III. Gaseous C-H-O-N Chemistry with 9 Molecules}

\author{Kevin Heng\altaffilmark{1}}
\author{Shang-Min Tsai\altaffilmark{1}}
\altaffiltext{1}{University of Bern, Center for Space and Habitability, Sidlerstrasse 5, CH-3012, Bern, Switzerland.  Emails: kevin.heng@csh.unibe.ch, shang-min.tsai@space.unibe.ch}

\begin{abstract}
We present novel, analytical, equilibrium-chemistry formulae for the abundances of molecules in hot exoplanetary atmospheres that include the carbon, oxygen and nitrogen networks.  Our hydrogen-dominated solutions involve acetylene (C$_2$H$_2$), ammonia (NH$_3$), carbon dioxide (CO$_2$), carbon monoxide (CO), ethylene (C$_2$H$_4$), hydrogen cyanide (HCN), methane (CH$_4$), molecular nitrogen (N$_2$) and water (H$_2$O).  By considering only the gas phase, we prove that the mixing ratio of carbon monoxide is governed by a decic equation (polynomial equation of degree 10).  We validate our solutions against numerical calculations of equilibrium chemistry that perform Gibbs free energy minimization and demonstrate that they are accurate at the $\sim 1\%$ level for temperatures from 500--3000 K.  In hydrogen-dominated atmospheres, the ratio of abundances of HCN to CH$_4$ is nearly constant across a wide range of carbon-to-oxygen ratios, which makes it a robust diagnostic of the metallicity in the gas phase.  Our validated formulae allow for the convenient benchmarking of chemical kinetics codes and provide an efficient way of enforcing chemical equilibrium in atmospheric retrieval calculations.
\end{abstract}


\section{Introduction}

Atmospheric chemistry is an indispensible ingredient in the study of exoplanetary atmospheres, as it teaches the practitioner how and when to be surprised.  For example, if hydrogen-dominated atmospheres are in chemical equilibrium, then we expect the dominant carbon carriers to be methane and carbon monoxide at low and high temperatures, respectively.  To date, the contributions to the atmospheric chemistry literature have mostly taken the form of numerical calculations using equilibrium chemistry and chemical kinetics codes (e.g., \citealt{bs99,lodders02,line11,vm11,moses11,moses13a,moses13b,koppa12,madhu12,agundez14,hu14,hu15,venot15}).  A complementary approach, which is standard in the astrophysical literature, is to develop analytical models \citep{hlt16,hl16}.  The current study is the third in a series of papers devoted to constructing analytical models for exoplanetary atmospheres to aid in the development of intuition, following \cite{hw14} (for shallow-water fluid dynamics) and \cite{hml14} (for two-stream radiative transfer).

Specifically, \cite{hlt16} and \cite{hl16} have previously derived solutions for the relative abundances of molecules for purely gaseous chemistry and in C-H-O (carbon-hydrogen-oxygen) systems.  Here, we present novel, generalized analytical solutions for purely gaseous C-H-O-N (carbon-hydrogen-oxygen-nitrogen) systems with 6 and 9 molecules.  In \S\ref{sect:theory}, we concisely describe the theoretical setup, which we use to consider 6 and 9 molecules in \S\ref{sect:6} and \ref{sect:9}, respectively.  We present our results are in \S\ref{sect:results} and conclude in \S\ref{sect:conclusions}.  A \texttt{Python} script that implements our analytical formulae may be found at \texttt{http://github.com/exoclime/VULCAN}.

\section{Theoretical Preamble \& Setup}
\label{sect:theory}

\subsection{Net Chemical Reactions}

Our C-H-O-N network contains 6 net reactions,
\begin{equation}
\begin{split}
\mbox{CH}_4 + \mbox{H}_2\mbox{O} &\leftrightarrows \mbox{CO} + 3 \mbox{H}_2, \\
\mbox{CO}_2 + \mbox{H}_2 &\leftrightarrows \mbox{CO} + \mbox{H}_2\mbox{O}, \\
2\mbox{CH}_4 &\leftrightarrows \mbox{C}_2\mbox{H}_2 + 3 \mbox{H}_2, \\
\mbox{C}_2\mbox{H}_4 &\leftrightarrows \mbox{C}_2\mbox{H}_2 + \mbox{H}_2, \\
2 \mbox{NH}_3 &\leftrightarrows \mbox{N}_2 + 3 \mbox{H}_2, \\
\mbox{NH}_3 + \mbox{CH}_4 &\leftrightarrows  \mbox{HCN} + 3\mbox{H}_2.
\end{split}
\label{eq:reactions}
\end{equation}
The last three reactions, involving nitrogen, were informed by \cite{bs99}, \cite{lodders02} and \cite{moses11}.  Excluding molecular hydrogen, there are 9 molecules in total.

\subsection{Normalized Equilibrium Constants}

For each net reaction, there is a corresponding equilibrium constant.  In a departure from \cite{hlt16} and \cite{hl16}, we write the normalized equilibrium constants without primes as superscripts.  Obeying the order in equation (\ref{eq:reactions}), they are
\begin{equation}
\begin{split}
K &= \frac{\tilde{n}_{\rm CO}}{\tilde{n}_{\rm CH_4} \tilde{n}_{\rm H_2O}} = \left( \frac{P_0}{P} \right)^2 \exp{\left(-\frac{\Delta \tilde{G}_{0,1}}{{\cal R}_{\rm univ} T} \right)}, \\
K_2 &= \frac{\tilde{n}_{\rm CO} \tilde{n}_{\rm H_2O}}{\tilde{n}_{\rm CO_2}} = \exp{\left(-\frac{\Delta \tilde{G}_{0,2}}{{\cal R}_{\rm univ} T} \right)}, \\
K_3 &= \frac{\tilde{n}_{\rm C_2H_2}}{\tilde{n}_{\rm CH_4}^2} = \left( \frac{P_0}{P} \right)^2 \exp{\left(-\frac{\Delta \tilde{G}_{0,3}}{{\cal R}_{\rm univ} T} \right)}, \\
K_4 &= \frac{\tilde{n}_{\rm C_2H_2}}{\tilde{n}_{\rm C_2H_4}} = \frac{P_0}{P} \exp{\left(-\frac{\Delta \tilde{G}_{0,4}}{{\cal R}_{\rm univ} T} \right)}, \\
K_5 &= \frac{\tilde{n}_{\rm N_2}}{\tilde{n}_{\rm NH_3}^2} = \left( \frac{P_0}{P} \right)^2 \exp{\left(-\frac{\Delta \tilde{G}_{0,5}}{{\cal R}_{\rm univ} T} \right)}, \\
K_6 &= \frac{\tilde{n}_{\rm HCN}}{\tilde{n}_{\rm NH_3} \tilde{n}_{\rm CH_4}} = \left( \frac{P_0}{P} \right)^2 \exp{\left(-\frac{\Delta \tilde{G}_{0,6}}{{\cal R}_{\rm univ} T} \right)}, \\
\end{split}
\label{eq:eq_constants}
\end{equation}
where $P_0=1$ bar is the reference pressure and ${\cal R}_{\rm univ} = 8.3144621$ J K$^{-1}$ mol$^{-1}$ is the universal gas constant.  For a molecule X, we have defined $\tilde{n}_{\rm X} \equiv n_{\rm X}/n_{\rm H_2}$, where $n_{\rm X}$ denotes the number density.  We call $\tilde{n}_{\rm X}$ the ``mixing ratios".

Appendix \ref{append:gibbs} lists the Gibbs free energies of formation for C$_2$H$_4$, N$_2$, NH$_3$ and HCN.  Appendix \ref{append:gibbs} also lists the Gibbs free energies associated with the last three net reactions (denoted by $\Delta \tilde{G}_{0,i}$ for the $i$-th reaction).  The Gibbs free energies for the other molecules and the first three net reactions have previously been stated in Tables 1 and 2 of \cite{hl16}, respectively.

\begin{figure}
\begin{center}
\includegraphics[width=\columnwidth]{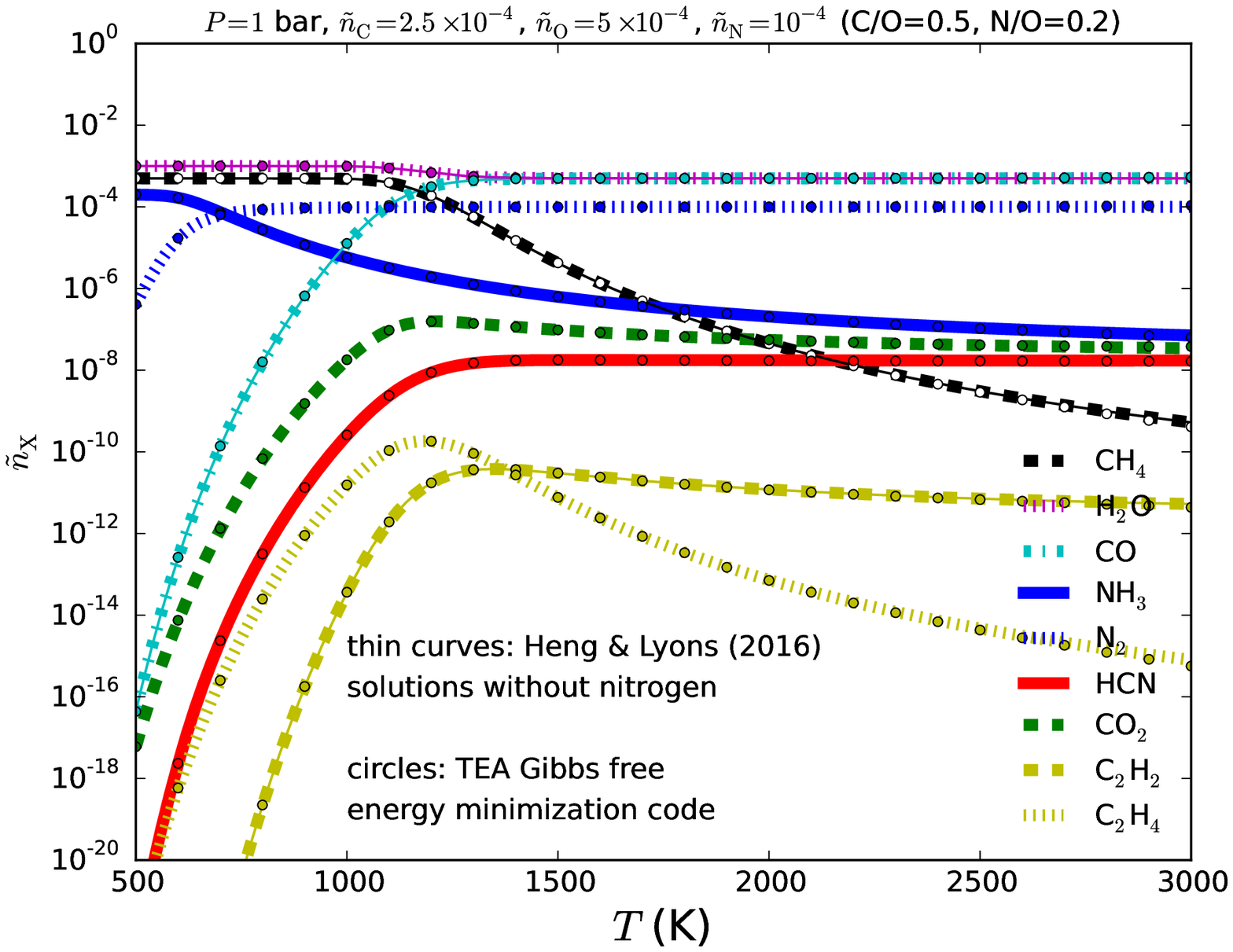}
\includegraphics[width=\columnwidth]{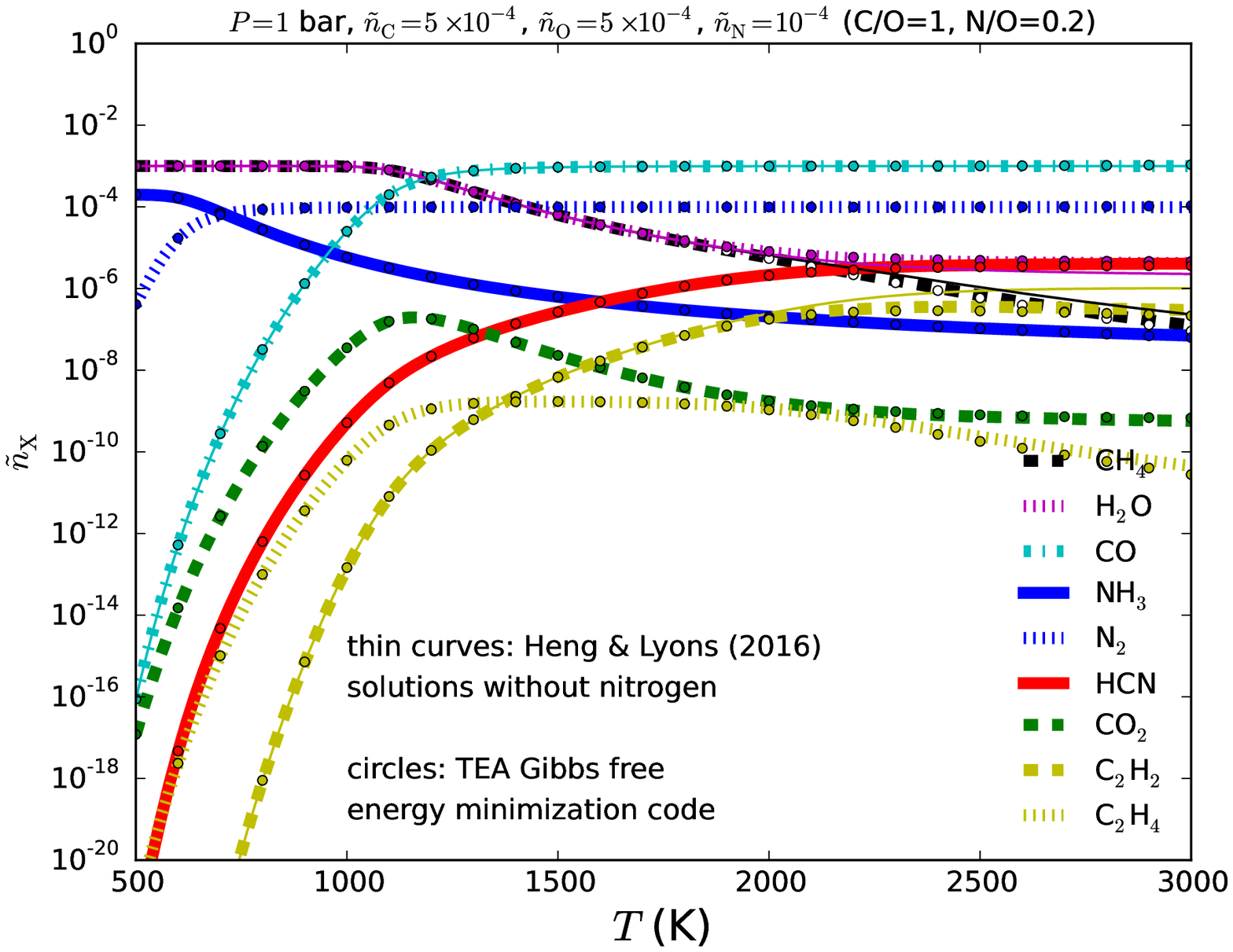}
\includegraphics[width=\columnwidth]{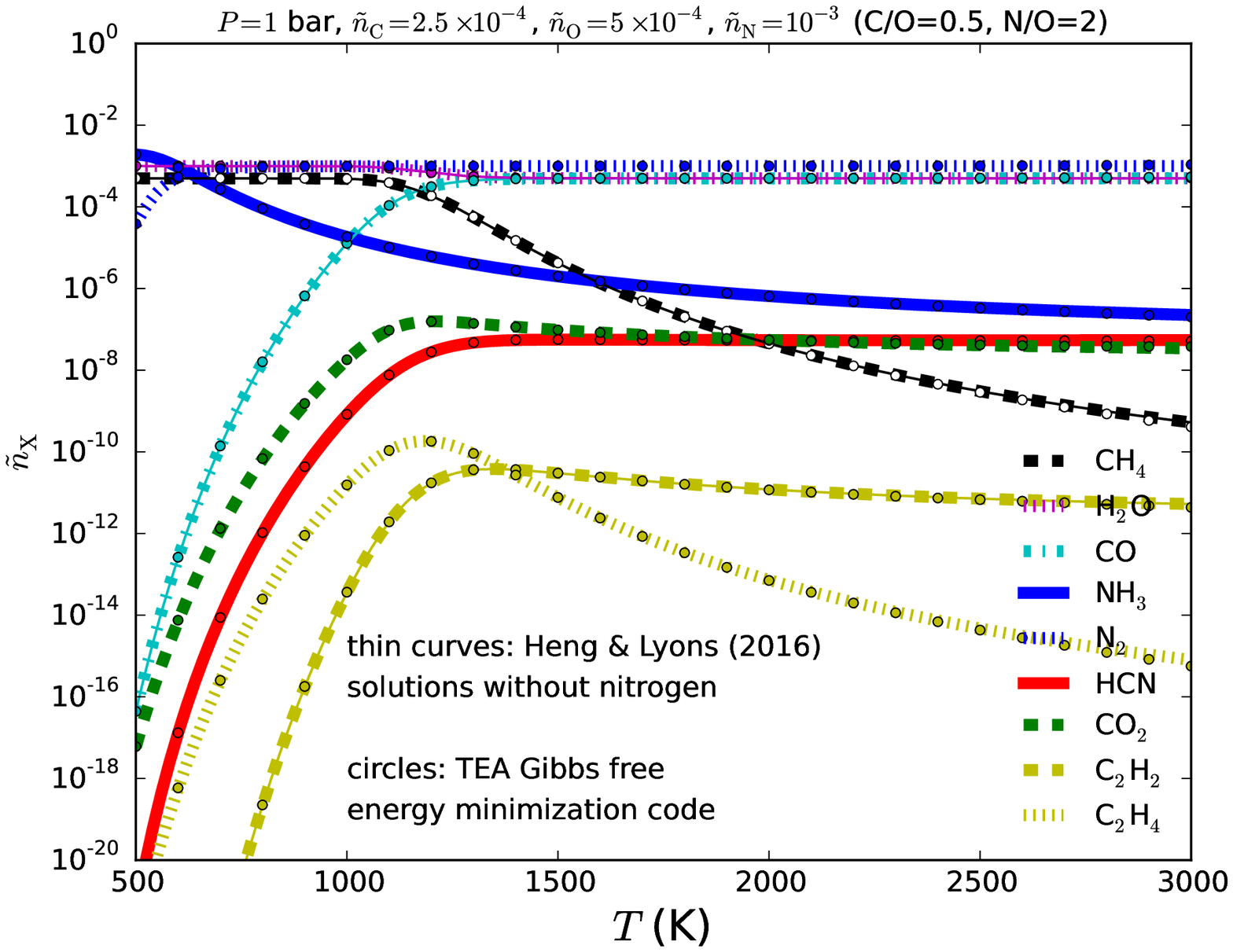}
\end{center}
\caption{Mixing ratios of various molecules versus temperature for solar-abundance (top panel), $\mbox{C/O}=1$ (middle panel) and nitrogen-rich (bottom panel) atmospheres.  For reference, we have plotted, as thin curves, the nitrogen-free solutions of \cite{hl16} for acetylene, carbon monoxide, methane and water.  The circles represent calculations done using the \texttt{TEA} code (see text).}
\label{fig:temperature}
\end{figure}

\subsection{Particle Conservation Equations}

By counting the number of atoms sequestered in each molecule, we obtain
\begin{equation}
\begin{split}
n_{\rm CH_4} &+ n_{\rm CO} + n_{\rm CO_2} + 2 n_{\rm C_2H_2} + 2 n_{\rm C_2H_4} \\
&+ n_{\rm HCN} = n_{\rm C}, \\
4n_{\rm CH_4} &+ 2 n_{\rm H_2O} + 2n_{\rm H_2} + 2 n_{\rm C_2H_2} + 4 n_{\rm C_2H_4} \\
&+ n_{\rm HCN} + 3 n_{\rm NH_3} = n_{\rm H}, \\
n_{\rm H_2O} &+ n_{\rm CO} + 2 n_{\rm CO_2} = n_{\rm O}, \\
2 n_{\rm N_2} &+ n_{\rm NH_3} + n_{\rm HCN} = n_{\rm N}.
\end{split}
\end{equation}
The number densities of atomic carbon, hydrogen, oxygen and nitrogen are given by $n_{\rm C}$, $n_{\rm H}$, $n_{\rm O}$ and $n_{\rm N}$, respectively.     

\section{C-H-O-N Network with 6 Molecules}
\label{sect:6}

\begin{figure}
\begin{center}
\vspace{0.2in}
\includegraphics[width=\columnwidth]{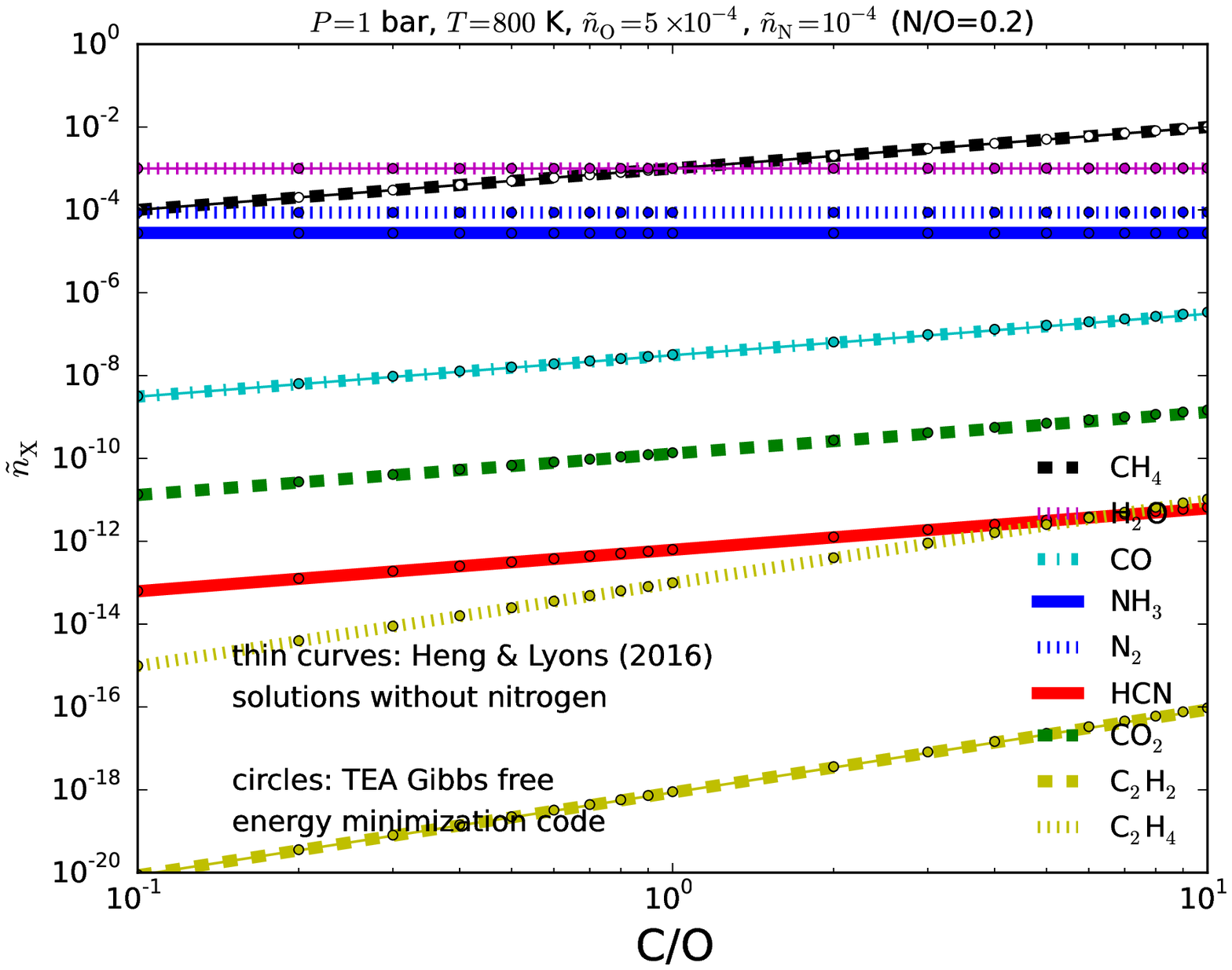}
\includegraphics[width=\columnwidth]{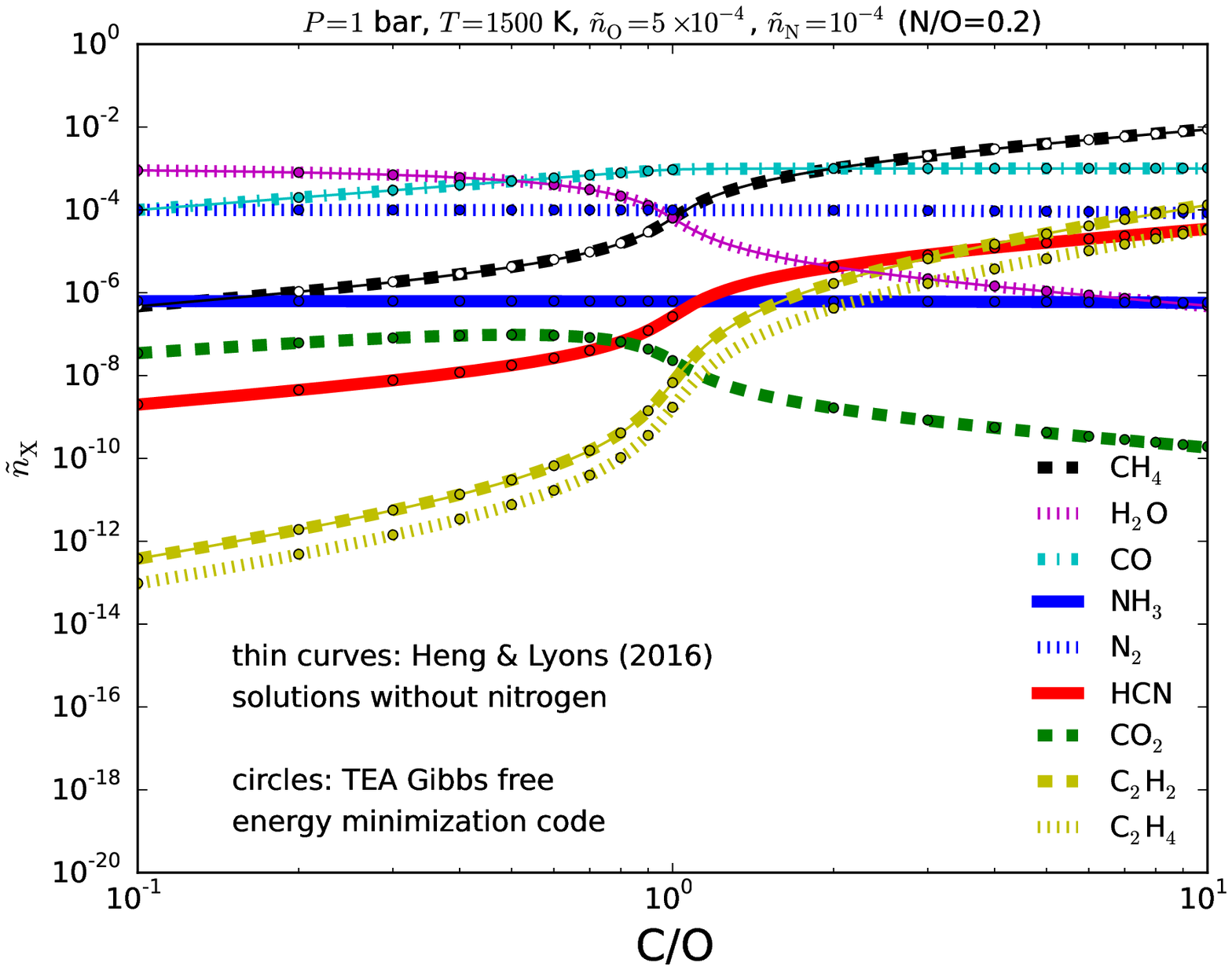}
\end{center}
\caption{Mixing ratios of various molecules versus the carbon-to-oxygen ratio.  Top panel: 800 K.  Bottom panel: 1500 K.  For reference, we have plotted, as thin curves, the nitrogen-free solutions of \cite{hl16} for acetylene, carbon monoxide, methane and water.  The circles represent calculations done using the \texttt{TEA} code (see text).}
\label{fig:co_ratio}
\end{figure}

To develop our current analytical method, we first ignore CO$_2$, C$_2$H$_2$ and C$_2$H$_4$ in our analysis.  \cite{hl16} have shown that CO$_2$ is subdominant compared to CO and H$_2$O, unless the metallicity is orders of magnitude higher than solar abundance.

If we make the simplification that $2n_{\rm H_2} = n_{\rm H}$ (hydrogen-dominated atmospheres) and render the number densities dimensionless, then we end up with
\begin{equation}
\begin{split}
\tilde{n}_{\rm CH_4} &+ \tilde{n}_{\rm CO} + \tilde{n}_{\rm HCN} = 2 \tilde{n}_{\rm C}, \\
\tilde{n}_{\rm H_2O} &+ \tilde{n}_{\rm CO} = 2 \tilde{n}_{\rm O}, \\
2 \tilde{n}_{\rm N_2} &+ \tilde{n}_{\rm NH_3} + \tilde{n}_{\rm HCN} = 2 \tilde{n}_{\rm N}.
\end{split}
\end{equation}
We define the elemental abundances as $\tilde{n}_{\rm C} \equiv n_{\rm C}/n_{\rm H}$, $\tilde{n}_{\rm O} \equiv n_{\rm O}/n_{\rm H}$ and $\tilde{n}_{\rm N} \equiv n_{\rm N}/n_{\rm H}$.  The goal is to decouple this system of non-linear equations such that one obtains a polynomial equation describing only one of the molecules.  This requires that we first rewrite some of the mixing ratios in terms of only $\tilde{n}_{\rm CO}$ and $\tilde{n}_{\rm H_2O}$,
\begin{equation}
\begin{split}
\tilde{n}_{\rm CH_4} &= \frac{\tilde{n}_{\rm CO}}{K \tilde{n}_{\rm H_2O}}, \\
\tilde{n}_{\rm NH_3} &= \frac{K \tilde{n}_{\rm CO}^2 - D_1 \tilde{n}_{\rm CO} + 4 K \tilde{n}_{\rm C} \tilde{n}_{\rm O}}{K_6 \tilde{n}_{\rm CO}}, \\
\tilde{n}_{\rm HCN} &= \frac{K_6 \tilde{n}_{\rm NH_3} \tilde{n}_{\rm CO}}{K \tilde{n}_{\rm H_2O}}, \\
\tilde{n}_{\rm N_2} &= K_5 \tilde{n}_{\rm NH_3}^2,
\end{split}
\label{eq:others}
\end{equation}
where we have defined 
\begin{equation}
D_1 \equiv 1 + 2K \left( \tilde{n}_{\rm C} + \tilde{n}_{\rm O} \right).
\end{equation}
We then use 
\begin{equation}
\tilde{n}_{\rm H_2O} = 2 \tilde{n}_{\rm O} - \tilde{n}_{\rm CO}
\end{equation}
to eliminate the mixing ratio of water.

By substituting these expressions into the equation involving molecular nitrogen, we obtain a quintic equation for the mixing ratio of CO,
\begin{equation}
\sum^5_{i=0} A_i \tilde{n}_{\rm CO}^i = 0.
\label{eq:quintic}
\end{equation}
The coefficients of this quintic equation are
\begin{equation}
\begin{split}
A_0 =& 256 K^3 K_5 \tilde{n}_{\rm O}^3 \tilde{n}_{\rm C}^2, \\
A_1 =& 32 K^2 \tilde{n}_{\rm O}^2 \tilde{n}_{\rm C} \left[ K_6 - 4 K_5 \left( D_1 + K \tilde{n}_{\rm C} \right) \right], \\ 
A_2 =& 16 K K_5 \tilde{n}_{\rm O} \left( 8 K^2 \tilde{n}_{\rm O} \tilde{n}_{\rm C} + D_1^2 + 4 K D_1 \tilde{n}_{\rm C} \right) \\
&+ 8 K K_6 \tilde{n}_{\rm O} \left[ 2K_6 \left( \tilde{n}_{\rm C} - \tilde{n}_{\rm N} \right) - 2 K \tilde{n}_{\rm C} - D_1 \right], \\
A_3 =& -8 K K_5 \left( 4 K D_1 \tilde{n}_{\rm O} + 8 K^2 \tilde{n}_{\rm O} \tilde{n}_{\rm C} + D_1^2 \right) \\
&+ 4 K K_6 \left( 2 K \tilde{n}_{\rm O} + D_1 \right) + 4 K_6^2 \left( 2 K \tilde{n}_{\rm N} - D_1 \right), \\
A_4 &= 16 K^2 K_5  \left( K \tilde{n}_{\rm O} + D_1 \right) + 4K K_6 \left( K_6 - K \right), \\
A_5 &= - 8 K^3 K_5.
\end{split}
\end{equation}
This derivation demonstrates that it is possible to decouple a C-H-O-N system and obtain an equation in terms of only $\tilde{n}_{\rm CO}$.

\section{C-H-O-N Network with 9 Molecules}
\label{sect:9}

We now add CO$_2$, C$_2$H$_2$ and C$_2$H$_4$ back into the analysis and use the method developed in the previous section.  The particle conservation equations become
\begin{equation}
\begin{split}
\tilde{n}_{\rm CH_4} &+ \tilde{n}_{\rm CO} + \tilde{n}_{\rm CO_2} + \tilde{n}_{\rm HCN} \\
&+ 2 \tilde{n}_{\rm C_2H_2} + 2 \tilde{n}_{\rm C_2H_4} = 2 \tilde{n}_{\rm C}, \\
\tilde{n}_{\rm H_2O} &+ \tilde{n}_{\rm CO} + 2 \tilde{n}_{\rm CO_2} = 2 \tilde{n}_{\rm O}, \\
2 \tilde{n}_{\rm N_2} &+ \tilde{n}_{\rm NH_3} + \tilde{n}_{\rm HCN} = 2 \tilde{n}_{\rm N}.
\end{split}
\end{equation}
The various mixing ratios are now described by
\begin{equation}
\begin{split}
\tilde{n}_{\rm H_2O} &= \frac{D_5}{D_4}, \\
\tilde{n}_{\rm CO_2} &= \frac{\tilde{n}_{\rm CO} \tilde{n}_{\rm H_2O}}{K_2}, \\
\tilde{n}_{\rm NH_3} &= \frac{D_2}{K K_6 D_4^2 D_5 \tilde{n}_{\rm CO}}, \\
\tilde{n}_{\rm HCN} &= \frac{K_6 D_4 \tilde{n}_{\rm NH_3} \tilde{n}_{\rm CO}}{K D_5}. \\
\end{split}
\label{eq:others2}
\end{equation}
The expressions for $\tilde{n}_{\rm CH_4}$ and $\tilde{n}_{\rm N_2}$ are the same as those given in equation (\ref{eq:others}).  To make the algebra tractable, we have defined
\begin{equation}
\begin{split}
D_2 \equiv& -K D^2_4 D_5 \tilde{n}_{\rm CO} + K^2 D_4 D_5^2 \left( 2 \tilde{n}_{\rm C} - \tilde{n}_{\rm CO} \right)  \\
&- 2 K_3 D_3 D_4^3 \tilde{n}_{\rm CO}^2 - \frac{K^2 D_5^3 \tilde{n}_{\rm CO}}{K_2}, \\
D_3 \equiv& 1 + \frac{1}{K_4}, \\
D_4 \equiv& 1 + \frac{2 \tilde{n}_{\rm CO}}{K_2}, \\
D_5 \equiv& 2 \tilde{n}_{\rm O} - \tilde{n}_{\rm CO}.
\end{split}
\end{equation}
If we write $D_2$ as
\begin{equation}
D_2 = \sum^5_{i=0} F_i \tilde{n}_{\rm CO}^i,
\end{equation}
then one may show that the coefficients are
\begin{equation}
\begin{split}
F_0 =& 8 K^2 \tilde{n}_{\rm O}^2 \tilde{n}_{\rm C}, \\
F_1 =& 2 K \tilde{n}_{\rm O} \left\{ -1 + 2K \left[ 2 \tilde{n}_{\rm C} \left( \frac{2 \tilde{n}_{\rm O}}{K_2} - 1 \right) - \tilde{n}_{\rm O} \right] \right.\\
&\left.- \frac{4K \tilde{n}_{\rm O}^2}{K_2} \right\}, \\
F_2 =& K \left( 1 - \frac{8 \tilde{n}_{\rm O}}{K_2} \right) - 2 K_3 D_3 \\
& + 2K^2 \left[ \tilde{n}_{\rm C} \left( 1 - \frac{8 \tilde{n}_{\rm O}}{K_2} \right) + 2 \tilde{n}_{\rm O} \left( 1 + \frac{\tilde{n}_{\rm O}}{K_2} \right) \right], \\
F_3 =& \frac{4K}{K_2} \left( 1 - \frac{2 \tilde{n}_{\rm O}}{K_2} \right) - \frac{12 K_3 D_3}{K_2} \\
&+ K^2 \left[ \frac{2\left(2 \tilde{n}_{\rm C} + \tilde{n}_{\rm O} \right)}{K_2} - 1 \right], \\
F_4 =& \frac{K}{K_2} \left( \frac{4}{K_2} - K \right) - \frac{24 K_3 D_3}{K^2_2} , \\
F_5 =& -\frac{16 K_3 D_3}{K^3_2}.
\end{split}
\label{eq:fcoefficients}
\end{equation}

Again, the goal is to obtain a single equation for $\tilde{n}_{\rm CO}$ by substituting all of these expressions into the equation involving molecular nitrogen,
\begin{equation}
\begin{split}
&2 K_5 D_2^2 + K K_6 D_4^2 D_5 D_2 \tilde{n}_{\rm CO} \\
&- 2K^2 K_6^2 D_4^4 D_5^2 \tilde{n}_{\rm N} \tilde{n}_{\rm CO}^2 + K_6^2 D_4^3 D_2 \tilde{n}_{\rm CO}^2 = 0.
\end{split}
\end{equation}
Evaluating $D_2^2$ is particularly tedious (see Appendix \ref{append:coeff}).  It is worth nothing that $D_1$ and $D_3$ are essentially numbers that only depend on the normalized equilibrium constants, while $D_2$, $D_4$ and $D_5$ are functions of $\tilde{n}_{\rm CO}$.  By ``opening up" these terms in the preceding expression, we may re-express it as
\begin{equation}
\sum^{10}_{i=0} A_i \tilde{n}_{\rm CO}^i = 0.
\label{eq:key_equation}
\end{equation}
The coefficients of this decic equation are given in Appendix \ref{append:coeff}.    

\section{Results}
\label{sect:results}

\begin{figure}
\begin{center}
\includegraphics[width=\columnwidth]{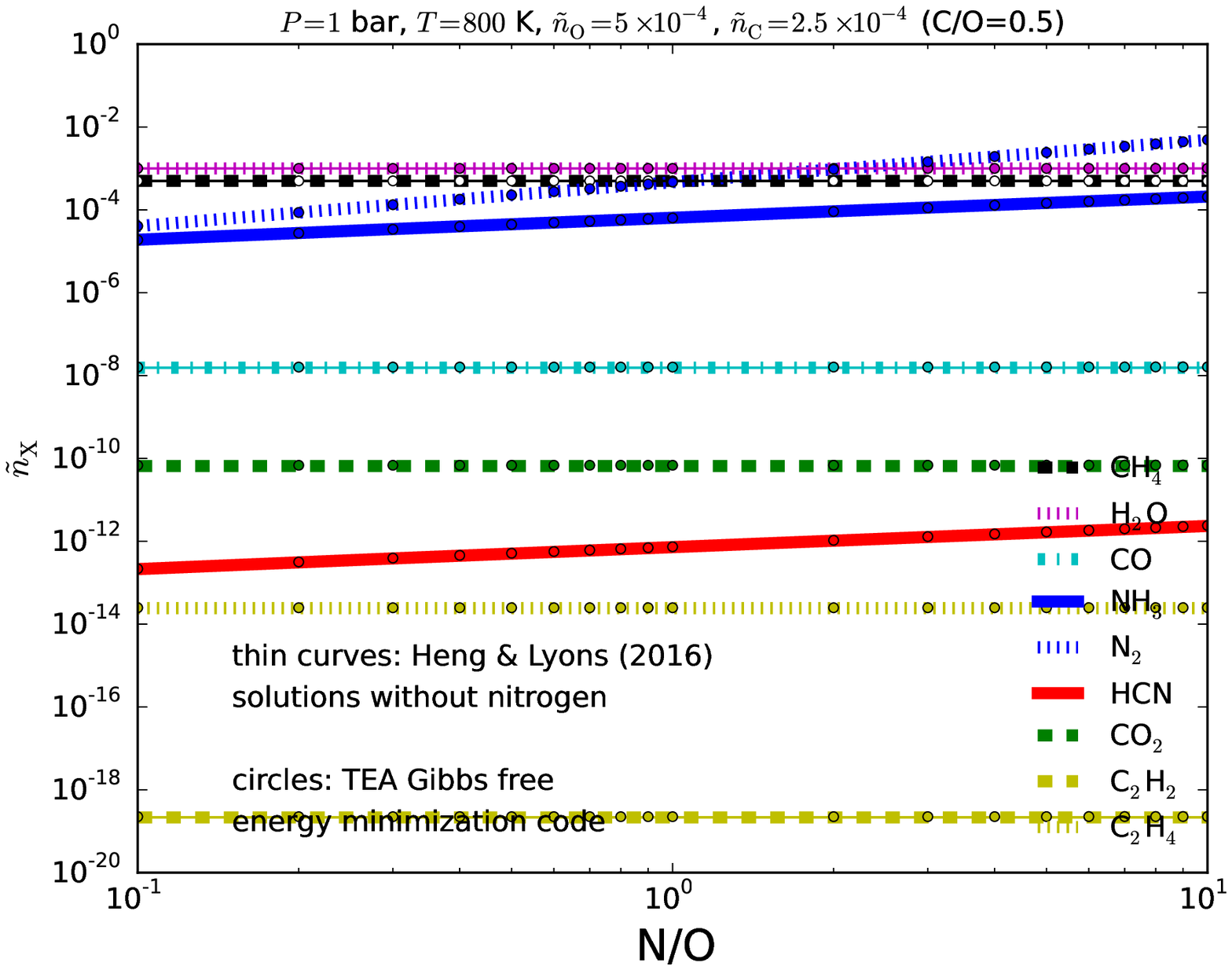}
\includegraphics[width=\columnwidth]{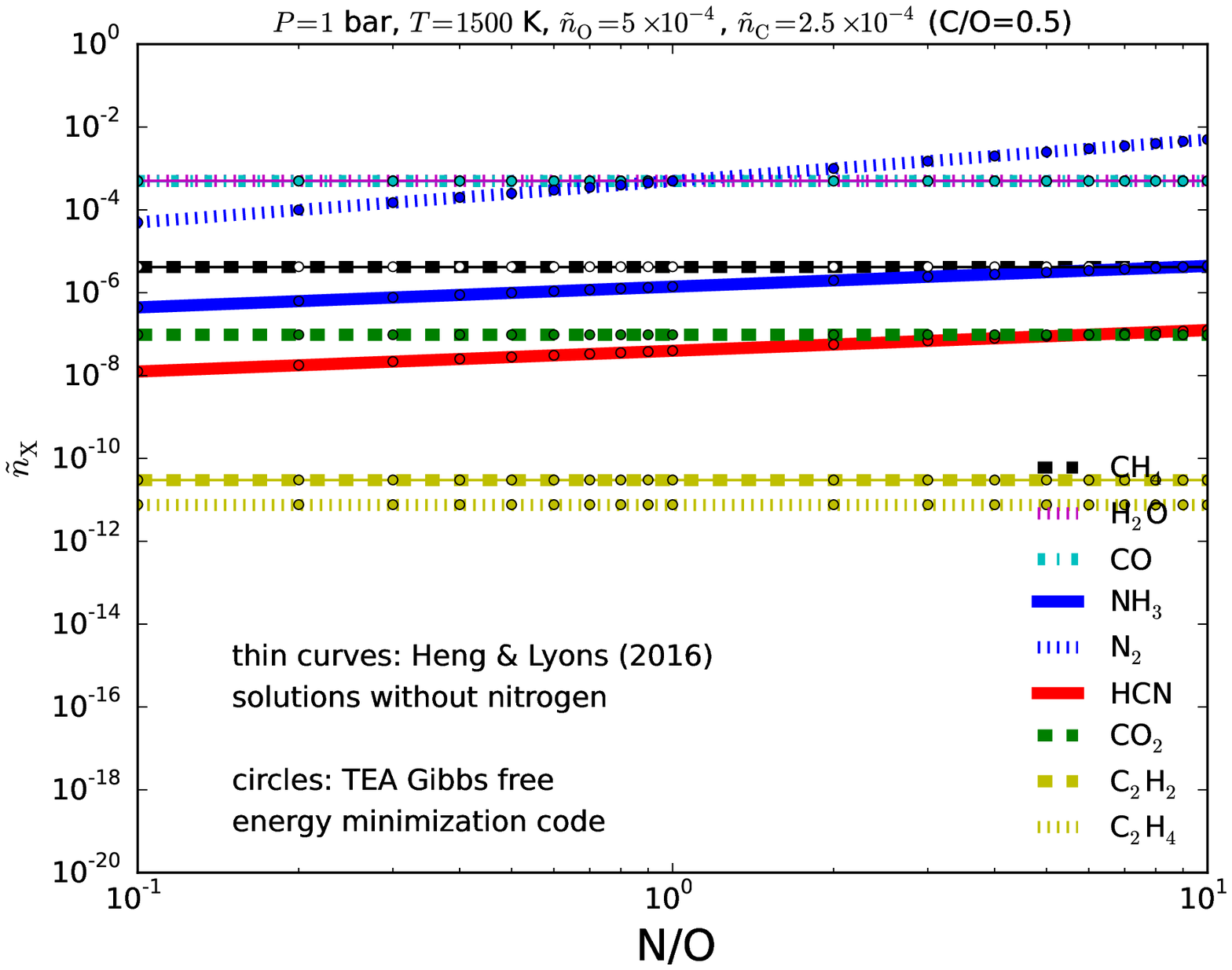}
\end{center}
\caption{Same as Figure \ref{fig:co_ratio}, but for trends versus the nitrogen-to-oxygen ratio.}
\label{fig:no_ratio}
\end{figure}

\begin{figure}
\begin{center}
\includegraphics[width=\columnwidth]{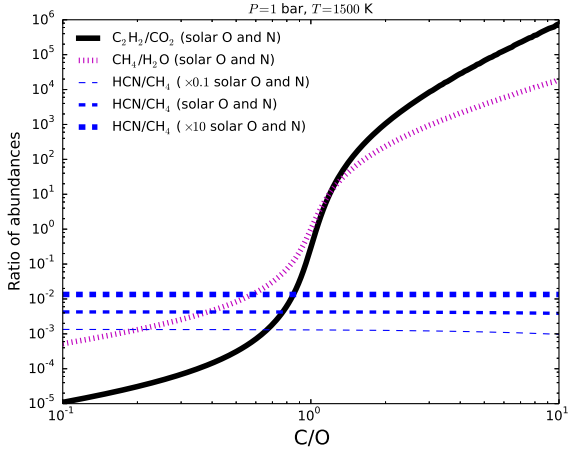}
\end{center}
\caption{Ratio of molecular abundances versus C/O.}
\label{fig:diagnostic}
\end{figure}

We define the solar abundance of elements as being $\tilde{n}_{\rm O} = 5 \times 10^{-4}$, $\tilde{n}_{\rm C} = 2.5 \times 10^{-4}$ and $\tilde{n}_{\rm N} = 10^{-4}$.  This implies that the carbon-to-oxygen and nitrogen-to-oxygen ratios are, respectively,
\begin{equation}
\mbox{C/O} \equiv \frac{\tilde{n}_{\rm C}}{\tilde{n}_{\rm O}} = 0.5 \mbox{ and } \mbox{N/O} \equiv \frac{\tilde{n}_{\rm N}}{\tilde{n}_{\rm O}} = 0.2.
\end{equation}
Unless otherwise stated, these are our default parameter values.

\subsection{Benchmarking}

We validate our analytical formulae by comparing them to calculations done using the \texttt{TEA} code \citep{blecic}, which performs Gibbs free energy minimization.  We emphasize that we use \texttt{TEA} at its full capability and include species beyond the 9 we have considered in our analytical formulae: H, C, O, H$_2$, O$_2$, CO, CO$_2$, CH$_4$, H$_2$O, CH, CH$_2$, CH$_3$, C$_2$H$_2$, OH, H$_2$CO, HCO, C$_2$, C$_2$H, C$_2$H$_4$, N$_2$, NH$_3$, HCN.  In attempting to solve the decic equation in (\ref{eq:key_equation}), we find that the procedure is sometimes numerically unstable, because the values of the various $A_i$ coefficients may vary by many orders of magnitude.  We emphasize that this is an issue of implementation and not of theory.  In practice, it is sufficient to obtain the mixing ratio of CO by solving the quintic equation in (\ref{eq:quintic}), which is numerically stable.  We use the \texttt{polyroots} routine in \texttt{Python}.  The other mixing ratios are then obtained using equations (\ref{eq:eq_constants}), (\ref{eq:others}) and (\ref{eq:others2}).

In Figures \ref{fig:temperature}, \ref{fig:co_ratio} and \ref{fig:no_ratio}, we represent calculations from the \texttt{TEA} code as circles overplotted on our calculations, which are shown as curves.  Using the \texttt{TEA} calculations as a reference, we calculate the errors associated with our analytical formulae.  For Figure \ref{fig:temperature}, we find that the errors are $\sim 1\%$, except for C$_2$H$_2$ and C$_2$H$_4$ with $\mbox{C/O}=1$ (middle panel) where they are $\sim 10\%$, but the increased inaccuracy is due to our use of the quintic, rather than the decic, equation.  For Figures \ref{fig:co_ratio} and \ref{fig:no_ratio}, we find that the errors are $\sim 1\%$ or smaller.

\subsection{Basic Trends}

Figure \ref{fig:temperature} shows the trends associated with the mixing ratios versus temperature for solar-abundance, $\mbox{C/O}=1$ and nitrogen-rich atmospheres.  The trend of CH$_4$ and CO being the dominant carbon carriers at low and high temperatures, respectively, persists even in the presence of nitrogen \citep{madhu12}.  Analogously, NH$_3$ and N$_2$ are the dominant nitrogen carriers at low and high temperatures, respectively \citep{bs99,lodders02,moses11}.  HCN closely tracks the rise of CO with temperature.  In a $\mbox{C/O}=1$ environment, HCN inhibits the formation of CH$_4$ and C$_2$H$_2$, as we can see from comparing the solutions derived in the current study versus the nitrogen-free solutions of \cite{hl16}.  Even when N/O is increased tenfold from 0.2 to 2, the trends produced resemble those of the solar-abundance case.

Figure \ref{fig:co_ratio} shows the mixing ratios versus C/O.  The low- and high-temperature trends have previously been elucidated, namely that carbon-rich atmospheres are water-poor and methane-rich \citep{madhu12,moses13a,hlt16}.  HCN closely tracks the abundance of CH$_4$ as C/O increases, suggesting that the ratio of their abundances should be a constant.  Figure \ref{fig:no_ratio} shows that the mixing ratios are somewhat insensitive to N/O.  Unsurprisingly, only the nitrogen-bearing species (N$_2$, NH$_3$ and HCN) show any dependence of their mixing ratios on N/O.

We note that our formulae do not consider graphite formation, which is expected to occur in carbon-rich atmospheres \citep{moses13b}.  For this reason, we urge caution when applying these formulae to $\mbox{C/O}>1$ situations.

\subsection{Observational Diagnostics}

Figure \ref{fig:diagnostic} shows the ratio of abundances of various pairs of molecules.  The C$_2$H$_2$/CO$_2$ ratio is a sensitive diagnostic for C/O \citep{venot15,hl16}, spanning more than 10 orders of magnitude as C/O varies from 0.1 to 10, suggesting that this ratio may be used as an observational diagnostic for inferring the value of C/O, provided a given spectrum of an exoplanetary atmosphere has the sufficient resolution and signal-to-noise for such an inference to be made via an inversion technique.  The CH$_4$/H$_2$O ratio is somewhat less sensitive to C/O, but provides an additional check on the inferred value of C/O.  The HCN/CH$_4$ ratio is essentially constant across a factor of 100 in C/O and its value depends only on the metallicity, implying that it may be used as a robust diagnostic for the metallicity of the atmosphere.  

\section{Conclusions \& Implications}
\label{sect:conclusions}

We have developed a novel analytical method for computing the abundances of 6 and 9 molecules in a C-H-O-N system in chemical equilibrium.  Our work demonstrates a useful trick, which is that trace molecules may formally be left out of the system of non-linear equations and computed later using the mixing ratios of other molecules.  Since our formulae have been successfully validated by a Gibbs free energy minimization code, they may be used to benchmark chemical kinetics codes.  Reproducing chemical equilibrium is a key test of a chemical kinetics code.  Our formulae may also be used in retrieval calculations to enforce chemical equilibrium throughout the atmosphere.  Such an approach may be used to test the Bayesian evidence for chemical disequilibrium when interpreting the spectrum of an exoplanetary atmosphere.

\acknowledgments
We acknowledge financial and administrative support from the Center for Space and Habitability (CSH), the PlanetS NCCR framework and the Swiss-based MERAC Foundation.  

\appendix

\section{Gibbs Free Energies of Molecules and Net Reactions}
\label{append:gibbs}

All data have been compiled using the NIST-JANAF database (\texttt{http://kinetics.nist.gov/janaf/}).  Note that the molar Gibbs free energy associated with N$_2$ and H$_2$ are 0 J mol$^{-1}$ by definition.  The Gibbs free energy of formation for C$_2$H$_4$ are (in units of kJ/mol and from 500 to 3000 K, in intervals of 100 K): 80.933, 88.017, 95.467, 103.180, 111.082, 119.122, 127.259, 135.467, 143.724, 152.016, 160.331, 168.663, 177.007, 185.357, 193.712, 202.070, 210.429, 218.790, 227.152, 235.515, 243.880, 252.246, 260.615, 268.987, 277.363, 285.743.  For NH$_3$, we have: 4.800, 15.879, 27.190, 38.662, 50.247, 61.910, 73.625, 85.373, 97.141, 108.918, 120.696, 132.469, 144.234, 155.986, 167.725, 179.447, 191.152, 202.840, 214.509, 226.160, 237.792, 249.406, 261.003, 272.581, 284.143, 295.689.  For HCN, we have: 117.769, 114.393, 111.063, 107.775, 104.525, 101.308, 98.120, 94.955, 91.812, 88.687, 85.579, 82.484, 79.403, 76.333, 73.274, 70.226, 67.187, 64.158, 61.138, 58.127, 55.124, 52.130, 49.144, 46.167, 43.198, 40.237.  For $\Delta \tilde{G}_{0,4}$, we have: 116.519, 103.718, 90.63, 77.354, 63.959, 50.485, 36.967, 23.421, 9.864, -3.697, -17.253, -30.802, -44.342, -57.87, -71.385, -84.888, -98.377, -111.855, -125.322, -138.777, -152.222, -165.657, -179.085, -192.504, -205.916, -219.322.  For $\Delta \tilde{G}_{0,5}$, we have: -9.6, -31.758, -54.38, -77.324, -100.494, -123.82, -147.25, -170.746, -194.282, -217.836, -241.392, -264.938, -288.468, -311.972, -335.45, -358.894, -382.304, -405.68, -429.018, -452.32, -475.584, -498.812, -522.006, -545.162, -568.286, -591.378.  For $\Delta \tilde{G}_{0,6}$, we have: 145.71, 121.401, 96.516, 71.228, 45.662, 19.906, -5.977, -31.942, -57.955, -83.992, -110.035, -136.073, -162.096, -188.098, -214.075, -240.023, -265.94, -291.826, -317.679, -343.5, -369.29, -395.047, -420.775, -446.472, -472.141, -497.784.

\section{System with 8 Molecules}

If only CO$_2$ is excluded, then we end up with a hexic/sextic equation for $\tilde{n}_{\rm CO}$ with the following coefficients,
\begin{equation}
\begin{split}
A_0 =& 2 K_5 J_0, \\
A_1 =& 2 K_5 J_1 + 2 K K_6 \tilde{n}_{\rm O} F_0, \\
A_2 =& 2 K_5 J_2 + K K_6 \left( 2 \tilde{n}_{\rm O} F_1 - F_0 \right) + K^2_6 F_0 - 8 K^2 K^2_6 \tilde{n}_{\rm O}^2 \tilde{n}_{\rm N}, \\
A_3 =& 2 K_5 J_3 + K K_6 \left( 2 \tilde{n}_{\rm O} F_2 - F_1 \right) + K^2_6 F_1 + 8 K^2 K^2_6 \tilde{n}_{\rm O} \tilde{n}_{\rm N}, \\
A_4 =& 2 K_5 J_4 + K K_6 \left( 2 \tilde{n}_{\rm O} F_3 - F_2 \right) + K^2_6 F_2 - 2 K^2 K^2_6 \tilde{n}_{\rm N},\\
A_5 =& 2 K_5 J_5 + K_6 F_3 \left( K_6 - K \right), \\
A_6 =& 2 K_5 J_6,
\end{split}
\end{equation}
where $J_0 = F_0^2$, $J_1 = 2 F_0 F_1$, $J_2 = 2 F_0 F_2 + F_1^2$, $J_3 = 2 F_0 F_3 + 2 F_1 F_2$, $J_4 = 2 F_1 F_3 + F_2^2$, $J_5 = 2 F_2 F_3$, $J_6 = F_3^2$, $F_0 = 8 K^2 \tilde{n}_{\rm O}^2 \tilde{n}_{\rm C}$, $F_1 = - 2 K \tilde{n}_{\rm O} \left[ 1 + 2K \left( 2 \tilde{n}_{\rm C} + \tilde{n}_{\rm O} \right) \right]$, $F_2 = K - 2 K_3 D_3 + 2K^2 \left( \tilde{n}_{\rm C} + 2 \tilde{n}_{\rm O} \right)$ and $F_3 = - K^2$.

\section{Coefficients of Decic Equation for CO}
\label{append:coeff}

To render the algebra tractable, we have written
\begin{equation}
D_2^2 = \sum^{10}_{i=0} J_i \tilde{n}_{\rm CO}^i,
\end{equation}
where the coefficients are $J_0 = F_0^2$, $J_1 = 2 F_0 F_1$, $J_2 = F^2_1 + 2 F_0 F_2$, $J_3 = 2 F_0 F_3 + 2 F_1 F_2$, $J_4 = 2 F_0 F_4 + 2 F_1 F_3 + F_2^2$, $J_5 = 2 F_0 F_5 + 2 F_1 F_4 + 2 F_2 F_3$, $J_6 = 2 F_1 F_5 + 2 F_2 F_4 + F_3^2$, $J_7 = 2 F_2 F_5 + 2 F_3 F_4$, $J_8 = 2 F_3 F_5 + F^2_4$, $J_9 = 2 F_4 F_5$ and $J_{10} = F^2_5$.  The $F_i$ coefficients are defined in equation (\ref{eq:fcoefficients}).  For convenience, we also write $C_2 \equiv 1/K_2$.  The coefficients of equation (\ref{eq:key_equation}) are:
\begin{equation}
\begin{split}
A_0 =& 2 K_5 J_0, \\
A_1 =& 2 K_5 J_1 + 2 K K_6 \tilde{n}_{\rm O} F_0, \\
A_2 =& 2 K_5 J_2 + K K_6 \left[ 2 \tilde{n}_{\rm O} F_1 + F_0 \left( 8 C_2 \tilde{n}_{\rm O} - 1 \right) \right] + K^2_6 F_0 - 8 K^2 K^2_6 \tilde{n}_{\rm O}^2 \tilde{n}_{\rm N}, \\
A_3 =& 2 K_5 J_3 + K K_6 \left[ 2 \tilde{n}_{\rm O} F_2 + F_1 \left( 8 C_2 \tilde{n}_{\rm O} - 1 \right) + 4 F_0 C_2 \left( 2 C_2 \tilde{n}_{\rm O} - 1 \right) \right] + K^2_6 \left( F_1 + 6 F_0 C_2 \right) \\
&+ 8 K^2 K^2_6 \tilde{n}_{\rm O} \tilde{n}_{\rm N} \left( 1 - 8 C_2 \tilde{n}_{\rm O} \right), \\
A_4 =& 2 K_5 J_4 + K K_6 \left[ 2 \tilde{n}_{\rm O} F_3 + F_2 \left( 8 C_2 \tilde{n}_{\rm O} - 1 \right) + 4 F_1 C_2 \left( 2 C_2 \tilde{n}_{\rm O} - 1 \right) - 4 F_0 C^2_2 \right] \\
&+ K^2_6 \left( F_2 + 6 F_1 C_2 + 12 F_0 C^2_2 \right) - 2 K^2 K^2_6 \tilde{n}_{\rm N} \left( 1 - 32 C_2 \tilde{n}_{\rm O} + 96 C^2_2 \tilde{n}_{\rm O}^2 \right), \\
A_5 =& 2 K_5 J_5 + K K_6 \left[ 2 \tilde{n}_{\rm O} F_4 + F_3 \left( 8 C_2 \tilde{n}_{\rm O} - 1 \right) + 4 F_2 C_2 \left( 2 C_2 \tilde{n}_{\rm O} - 1 \right) - 4 F_1 C^2_2 \right] \\
&+ K^2_6 \left( F_3 + 6 F_2 C_2 + 12 F_1 C^2_2 + 8 F_0 C^3_2 \right) - 16 K^2 K^2_6 \tilde{n}_{\rm N} C_2 \left( 1 - 12 C_2 \tilde{n}_{\rm O} + 16 C^2_2 \tilde{n}_{\rm O}^2 \right), \\
A_6 =& 2 K_5 J_6 + K K_6 \left[ 2 \tilde{n}_{\rm O} F_5 + F_4 \left( 8 C_2 \tilde{n}_{\rm O} - 1 \right) + 4 F_3 C_2 \left( 2 C_2 \tilde{n}_{\rm O} - 1 \right) - 4 F_2 C^2_2 \right] \\
&+ K^2_6 \left( F_4 + 6 F_3 C_2 + 12 F_2 C^2_2 + 8 F_1 C^3_2 \right) - 16 K^2 K^2_6 \tilde{n}_{\rm N} C_2^2 \left( 3 - 16 C_2 \tilde{n}_{\rm O} + 8 C^2_2 \tilde{n}_{\rm O}^2 \right), \\
A_7 =& 2 K_5 J_7 + K K_6 \left[ F_5 \left( 8 C_2 \tilde{n}_{\rm O} - 1 \right) + 4 F_4 C_2 \left( 2 C_2 \tilde{n}_{\rm O} - 1 \right) - 4 F_3 C^2_2 \right] \\
&+ K^2_6 F_5 + 2 K^2_6 C_2 \left( 3 F_4 + 6 F_3 C_2 + 4 F_2 C^2_2 \right) - 64 K^2 K^2_6 \tilde{n}_{\rm N} C_2^3 \left( 1 - 2 C_2 \tilde{n}_{\rm O} \right), \\
A_8 =& 2 K_5 J_8 + 4 K K_6 C_2 \left[ F_5 \left( 2 C_2 \tilde{n}_{\rm O} - 1 \right) - F_4 C_2 \right] \\
&+ 2 K^2_6 C_2 \left( 3 F_5 + 6 F_4 C_2 + 4 F_3 C^2_2 \right) - 32 K^2 K^2_6 \tilde{n}_{\rm N} C_2^4, \\
A_9 =& 2 K_5 J_9 - 4 K K_6 F_5 C_2^2 + 4 K^2_6 C_2^2 \left( 3 F_5 + 2 F_4 C_2 \right), \\
A_{10} =& 2 K_5 J_{10} + 8 K^2_6 F_5 C_2^3.
\end{split}
\end{equation}
From a practical standpoint, if one was implementing these expressions in a computer code, one would first code up $K$ and $K_i$, followed by $F_i$ and $J_i$, which would allow the construction of $A_i$.

\section{Licensing and Permission to use the \texttt{TEA} Code}

We thank the developers of the Thermochemical Equilibrium Abundances (\texttt{TEA}) code \citep{blecic}, initially developed at the University of Central Florida, Orlando, Florida, USA.  The Reproducible Research Compendium (RRC) and the \texttt{Python} code we used to produce Figure \ref{fig:temperature} are available at \texttt{http://github.com/exoclime/VULCAN}.


\label{lastpage}

\end{document}